\documentclass[osajnl,twocolumn,showpacs,superscriptaddress,10pt]{revtex4-1} 
\usepackage{amsmath,amssymb,graphicx}
\usepackage{braket}
\begin{document}

\title{Absolute frequency measurements and hyperfine structures of the molecular iodine transitions at 578 nm}

\author{Takumi Kobayashi}\email{Corresponding author: takumi-kobayashi@aist.go.jp} 
\affiliation{National Metrology Institute of Japan (NMIJ), National Institute of Advanced Industrial Science and Technology (AIST), Tsukuba Central 3, Ibaraki 305-8563, Japan}
\affiliation{Department of Physics, Graduate School of Engineering, Yokohama National University, 79-5 Tokiwadai, Hodogaya-ku, Yokohama 240-8501, Japan}
\affiliation{JST, ERATO, MINOSHIMA Intelligent Optical Synthesizer Project, ERATO, 1-1-1 Umezono, Tsukuba, Ibaraki 305-8563, Japan}

\author{Daisuke Akamatsu}
\affiliation{National Metrology Institute of Japan (NMIJ), National Institute of Advanced Industrial Science and Technology (AIST), Tsukuba Central 3, Ibaraki 305-8563, Japan}

\author{Kazumoto Hosaka}
\author{Hajime Inaba}
\author{Sho Okubo}
\affiliation{National Metrology Institute of Japan (NMIJ), National Institute of Advanced Industrial Science and Technology (AIST), Tsukuba Central 3, Ibaraki 305-8563, Japan}
\affiliation{JST, ERATO, MINOSHIMA Intelligent Optical Synthesizer Project, ERATO, 1-1-1 Umezono, Tsukuba, Ibaraki 305-8563, Japan}

\author{Takehiko Tanabe}
\affiliation{National Metrology Institute of Japan (NMIJ), National Institute of Advanced Industrial Science and Technology (AIST), Tsukuba Central 3, Ibaraki 305-8563, Japan}

\author{Masami Yasuda}
\affiliation{National Metrology Institute of Japan (NMIJ), National Institute of Advanced Industrial Science and Technology (AIST), Tsukuba Central 3, Ibaraki 305-8563, Japan}

\author{Atsushi Onae}
\affiliation{National Metrology Institute of Japan (NMIJ), National Institute of Advanced Industrial Science and Technology (AIST), Tsukuba Central 3, Ibaraki 305-8563, Japan}
\affiliation{JST, ERATO, MINOSHIMA Intelligent Optical Synthesizer Project, ERATO, 1-1-1 Umezono, Tsukuba, Ibaraki 305-8563, Japan}
\author{Feng-Lei Hong}
\affiliation{Department of Physics, Graduate School of Engineering, Yokohama National University, 79-5 Tokiwadai, Hodogaya-ku, Yokohama 240-8501, Japan}
\affiliation{National Metrology Institute of Japan (NMIJ), National Institute of Advanced Industrial Science and Technology (AIST), Tsukuba Central 3, Ibaraki 305-8563, Japan}
\affiliation{JST, ERATO, MINOSHIMA Intelligent Optical Synthesizer Project, ERATO, 1-1-1 Umezono, Tsukuba, Ibaraki 305-8563, Japan}

\begin{abstract}We report absolute frequency measurements of 81 hyperfine components of the rovibrational transitions of molecular iodine at 578 nm using the second harmonic generation of an 1156-nm external-cavity diode laser and a fiber-based optical frequency comb. The relative uncertainties of the measured absolute frequencies are typically $1.4\times10^{-11}$. Accurate hyperfine constants of four rovibrational transitions are obtained by fitting the measured hyperfine splittings to a four-term effective Hamiltonian including the electric quadrupole, spin-rotation, tensor spin-spin, and scalar spin-spin interactions. The observed transitions can be good frequency references at 578 nm, and are especially useful for research using atomic ytterbium since the transitions are close to the $^{1}S_{0}$$-$$^{3}P_{0}$ clock transition of ytterbium. 
\end{abstract}


\maketitle 

\section{Introduction}
\label{introduction}
The precision spectroscopy of molecular iodine (I$_{2}$) is important for many applications, e.g., optical frequency metrology, optical communications, and studies in atomic and molecular physics. Many lasers that are stabilized to hyperfine components of the rovibrational transitions of I$_{2}$ are recommended as frequency standards by the Comit\'e International des Poids et Mesures (CIPM) \cite{Quinn2003,Felder2005}. Several groups have measured hyperfine components near 532 nm using a frequency-doubled Nd:YAG laser at 1064 nm with improving precision \cite{Arie1993,Eickhoff1995,Ye1999,Hong2000,Hong2004}, since I$_{2}$-stabilized Nd:YAG lasers at 532 nm have shown very high frequency stability \cite{Hall1999,Hong2004optcom,Zang2007}. Hyperfine components at 660 nm have been measured using a frequency-doubled Nd:YAG laser at 1319 nm \cite{Guo2004}. A frequency-doubled Nd:YAG laser at 1123 nm has been used to measure hyperfine components at 561 nm \cite{Yang2012}. More recently, a frequency-doubled diode laser at 1097 nm has been utilized to measure hyperfine components at 548 nm \cite{Hsiao2013}.

Iodine absorption lines at 578 nm are near the $^{1}S_{0}$$-$$^{3}P_{0}$ clock transition of atomic ytterbium (Yb) \cite{Hong2009}. Therefore, these iodine lines can be frequency references for Yb atom research, e.g., the development of an Yb optical lattice clock \cite{Kohno2009,Yasuda2012,Park2013,Hinkley2013}, and for some proposed experiments for quantum simulation \cite{Gorshkov2010,Gerbier2010} and quantum information processing \cite{Gorshkov2009,Daley2011}. These frequency references at 578 nm using a simple iodine cell are helpful as regards finding the very weak clock transition in Yb, especially for laboratories where no optical frequency combs are available. Frequency gaps between the clock transition and the iodine lines can be bridged using a high-finesse cavity \cite{Dareau2015} or sideband generation using an electro-optics modulator (EOM). In previous experiments \cite{Hong2009}, we have measured the absolute frequency of the $a_{1}$ hyperfine component of the $R(37)16-1$ rovibrational transition at 578 nm. Our measured frequency has recently been used for the spectroscopy of Bose-Einstein condensates of Yb using the clock transition \cite{Dareau2015}. To the best of our knowledge, however, there have been no reports on the absolute frequencies of other 578-nm lines of I$_{2}$. This paper presents absolute frequency measurements of 81 hyperfine components of four rovibrational transitions, namely the $R(37)16-1$, $P(33)16-1$, $R(101)17-1$, and $P(131)18-1$ transitions. 

\begin{figure*}[t]
\begin{center}
\includegraphics[width=12cm,bb=0 80 1010 716]{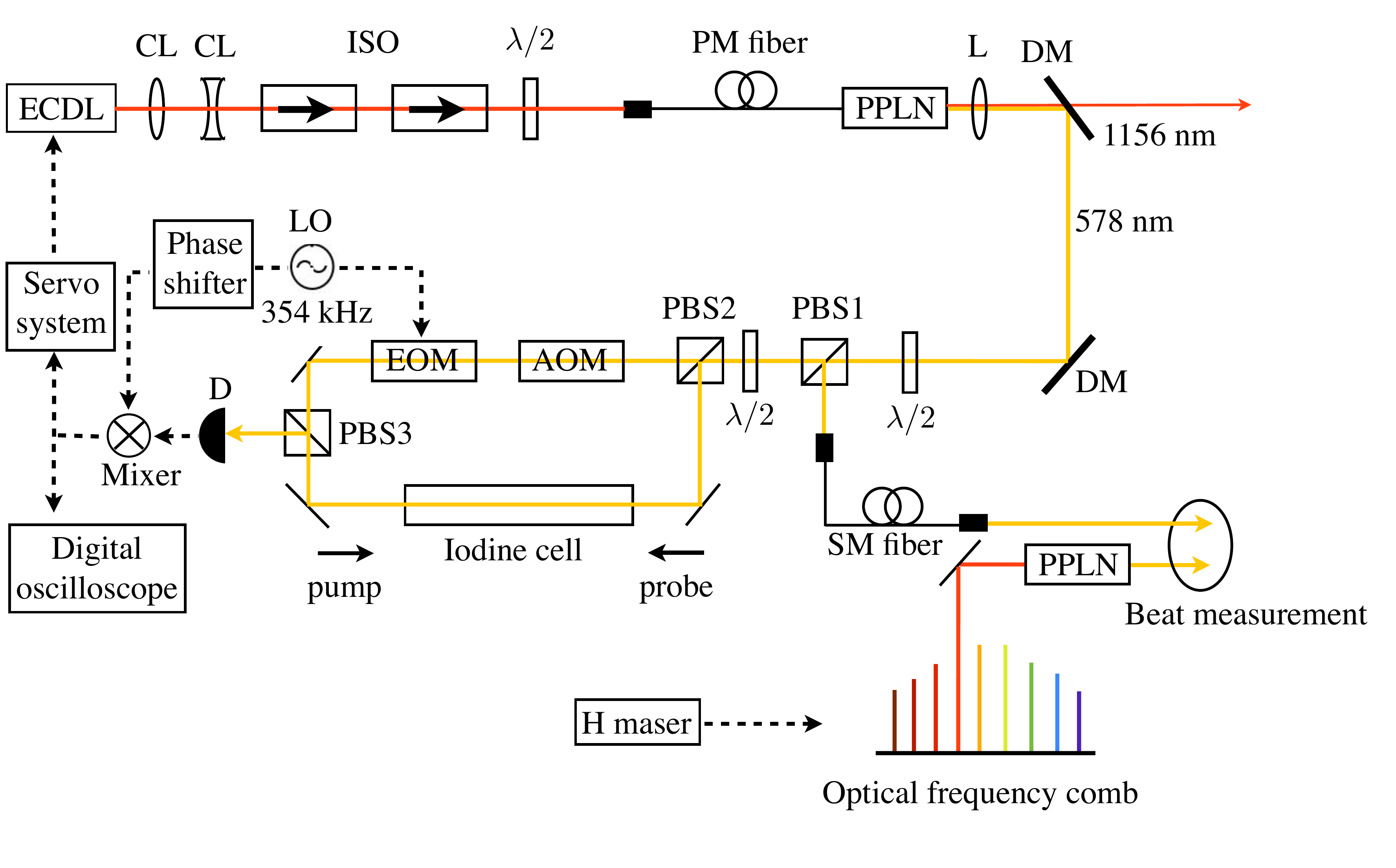}
\end{center}
\caption{Schematic diagram of the experimental setup. ECDL: External-cavity diode laser, CL: Cylindrical lens, ISO: Isolator, PM: Polarization-maintaining, PPLN: Periodically poled lithium niobate, L: Lens, DM: Dichroic mirror, PBS:  Polarization beam splitter, AOM: Acousto-optic modulator, EOM: Electro-optic modulator, D: Detector, LO: Local oscillator, $\lambda /2$: Half-wave plate.}
\label{figure1}
\end{figure*}

Since it is difficult to generate a yellow light near 578 nm directly using solid-state lasers or diode lasers, dye lasers have usually been used for spectroscopy experiments in this wavelength region \cite{Nishiyama2013}. As an alternative, we previously developed a 578-nm light source for an Yb optical lattice clock using the sum frequency generation (SFG) of a 1319-nm Nd:YAG laser and a 1030-nm Yb:YAG laser \cite{Hong2009,Hosaka2010}. Another way to generate a 578-nm light is to employ the second harmonic generation (SHG) of a diode laser at 1156 nm \cite{Nevsky2008,Kim2010,Lee2011}. The linewidth of a diode laser, e.g., an external-cavity diode laser (ECDL), is usually several hundred kilohertz, whereas that of the above SFG light source is several kilohertz \cite{Hong2009,Hosaka2010}. Nevertheless, the diode-laser-based SHG scheme is attractive due to its simplicity and compactness, and thus is employed in the present work. We investigate the frequency stability and the systematic frequency shifts of an I$_{2}$-stabilized SHG light source.

A theoretical fit of the measured hyperfine splittings of I$_{2}$ provides the hyperfine constants of I$_{2}$. The obtained hyperfine constants are important for improving or deriving formulas for the hyperfine constants of different rovibrational transitions. For example, the high frequency stability of an I$_{2}$-stabilized Nd:YAG laser near 532 nm has made it possible to perform precision measurements of hyperfine splittings and made an accurate determination of the hyperfine constants. These studies have revealed the rotation dependence of electric quadrupole hyperfine constants \cite{Hong2001JOSAB379,Hong2001JOSAB1416}, and the vibration dependence of tensor spin-spin and scalar spin-spin hyperfine constants \cite{Hong2002JOSAB}. Here we deduce the hyperfine constants of the observed four rovibrational transitions at 578 nm by employing a similar theoretical fit. 

\section{Experimental setup}
Figure \ref{figure1} is a schematic diagram of the experimental setup. The SHG light source at 578 nm consisted of a Littrow-type ECDL at 1156 nm using an InAs quantum dot laser (Innolume GmbH, GC-1156-TO-200) and a waveguide periodically poled lithium niobate (PPLN) crystal, which was manufactured by NTT Electronics. Two cylindrical lenses were used to circularize the output beam of the ECDL. This beam was allowed to pass through two optical isolators with an isolation of 76 dB and was then led into the PPLN via a polarization-maintaining (PM) fiber. This isolation was needed, since the reflected light from the fiber end made the laser frequency unstable. The generated SHG light at 578 nm with a power of 12 mW was collimated by a lens and separated from the fundamental light at 1156 nm using two dichroic mirrors. 

A half-wave plate ($\lambda/2$) and a polarization beam splitter (PBS1, see Fig. \ref{figure1}) were used to separate the 578-nm beam into two parts. A 2-3 mW portion of this beam was sent to an optical frequency comb via a single-mode (SM) fiber, while the other portion was sent to the sub-Doppler iodine spectrometer described below. The frequency comb was based on a mode-locked erbium-doped fiber laser operated at a repetition rate of 123 MHz \cite{Inaba2006,Nakajima2010,Iwakuni2012}, which was self-referenced \cite{Jones2000} and phase locked to a hydrogen maser (H maser). The frequency of the H-maser was calibrated against the Coordinated Universal Time of the National Metrology Institute of Japan (UTC(NMIJ)). A heterodyne beat note was detected between the 578-nm light and an SHG comb component, which was generated by passing the fundamental comb components through a waveguide PPLN. 

The sub-Doppler spectroscopy of I$_{2}$ was carried out based on the modulation transfer technique of saturation spectroscopy \cite{Eickhoff1995,Hong2009}. One of the major advantages of this technique is that it provides a nearly flat baseline of an observed spectrum. A $\lambda /2 $ plate and another polarization beam splitter (PBS2, see Fig. \ref{figure1}) were introduced to adjust the power ratio of the pump and probe beams. The pump beam was frequency shifted by 80 MHz using an acousto-optic modulator (AOM) and phase modulated by an EOM at a modulation frequency of 354 kHz. The AOM was introduced to prevent interferometric baseline problems in the iodine spectrometer. The pump and probe beams were overlapped anticollinearly in a 45-cm-long iodine cell. The powers of the pump and probe beams were both $P_{\mathrm{pump}}=4.8$ mW and $P_{\mathrm{probe}}=0.5$ mW, respectively. The diameters of the two beams inside the cell were $d\sim1.5$ mm. The cold-finger temperature of the iodine cell was held at $T=-2$ $^{\circ}$C, corresponding to an iodine pressure of $p=3.3$ Pa. The unmodulated probe beam developed sidebands inside the cell by a four-wave mixing process when saturation occurred. The probe beam was separated from the pump beam by a third polarization beam splitter (PBS3, see Fig. \ref{figure1}) and detected by a photo detector. The modulation transfer signal was obtained by demodulating the signal from the detector. This signal was fed back to the piezoelectric transducer (PZT) of the ECDL through a servo system for frequency stabilization. 

\section{Experimental results}
\subsection{Observation of hyperfine components}
\label{observationhyperfinesect}
\begin{figure}[h]
\begin{center}
\includegraphics[width=12cm,bb=0 80 1350 716]{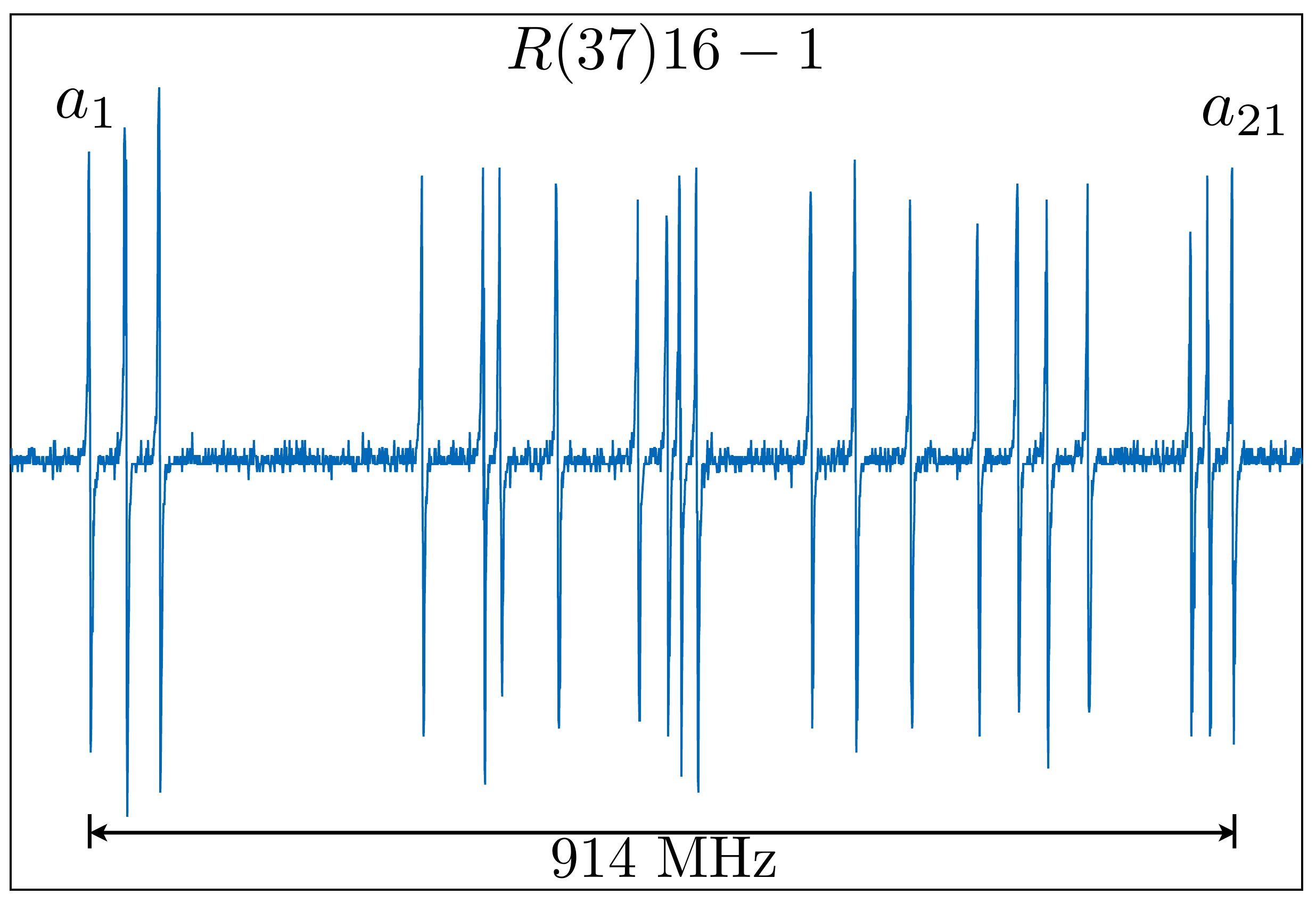}
\end{center}
\caption{Modulation transfer signals of the $R(37)16-1$ transition.}
\label{R37}
\end{figure}

\begin{figure}[h]
\begin{center}
\includegraphics[width=12cm,bb=0 80 1350 716]{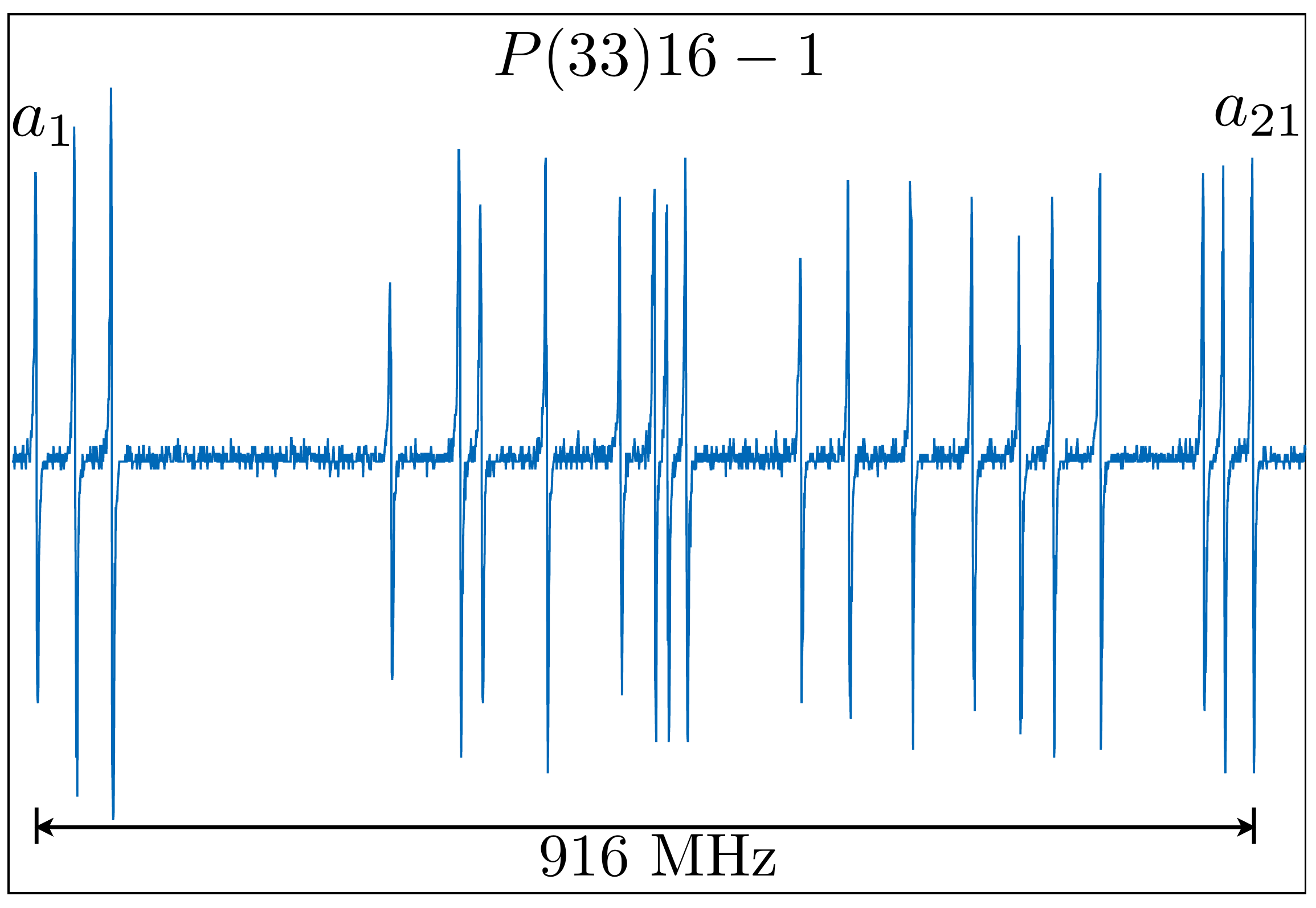}
\end{center}
\caption{Modulation transfer signals of the $P(33)16-1$ transition.}
\label{P33}
\end{figure}

\begin{figure}[h]
\begin{center}
\includegraphics[width=12cm,bb=0 80 1350 716]{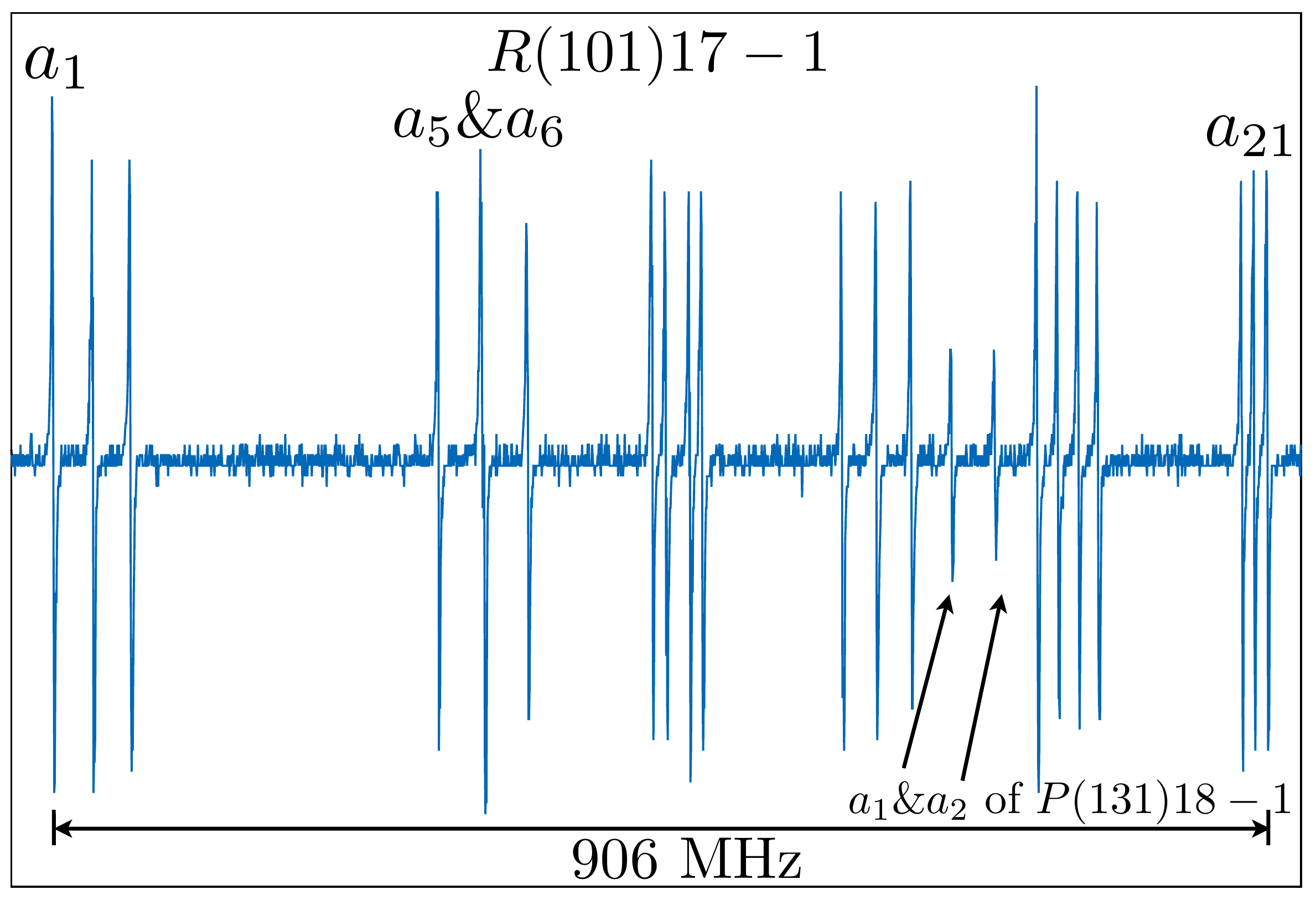}
\end{center}
\caption{Modulation transfer signals of the $R(101)17-1$ transition.}
\label{R101}
\end{figure}

\begin{figure}[h]
\begin{center}
\includegraphics[width=12cm,bb=0 80 1350 696]{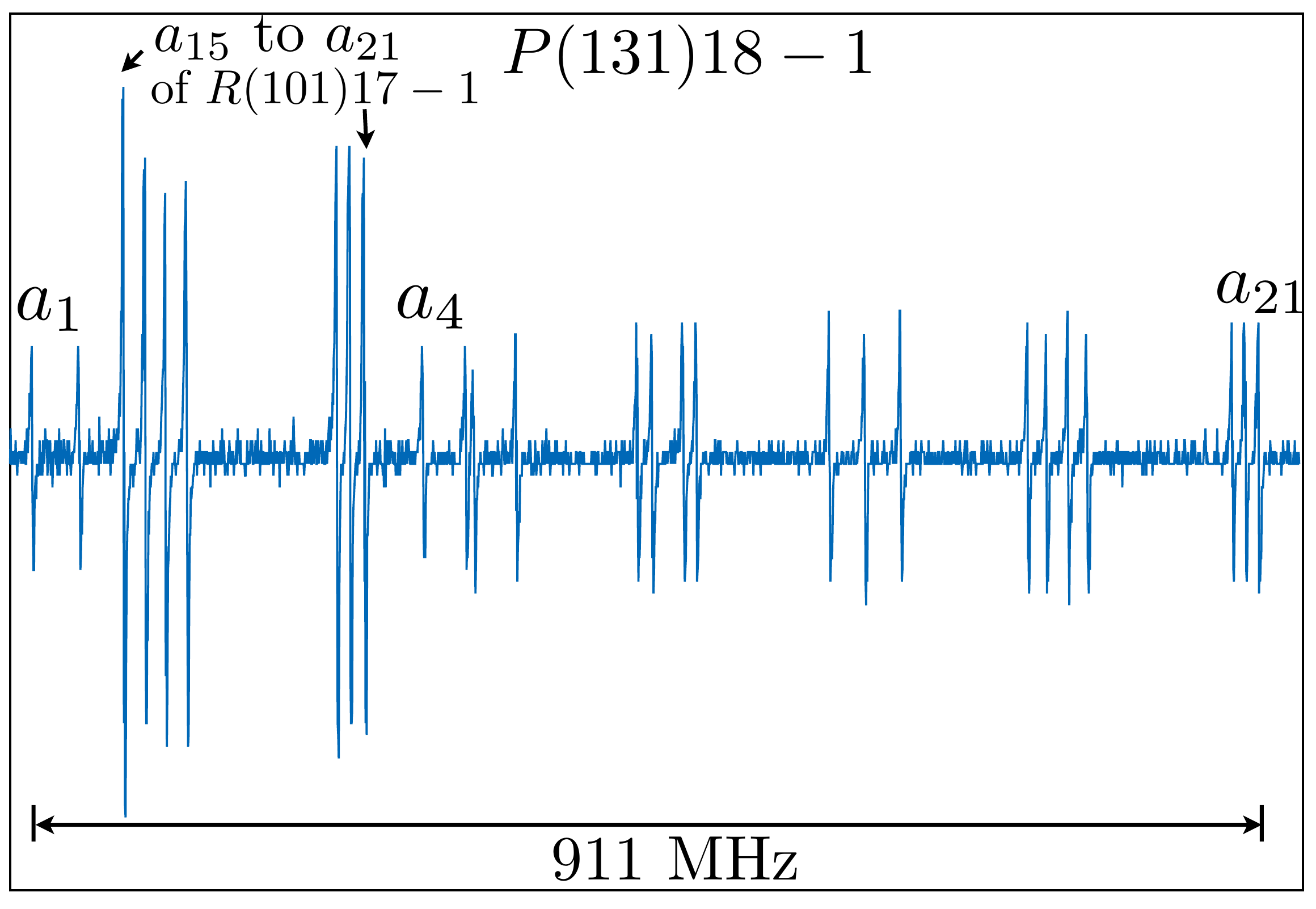}
\end{center}
\caption{Modulation transfer signals of the $P(131)18-1$ transition.}
\label{P131}
\end{figure}

Figures \ref{R37}, \ref{P33}, \ref{R101}, and \ref{P131} show the observed modulation transfer signals of the $R(37)16-1$, $P(33)16-1$, $R(101)17-1$, and $P(131)18-1$ transitions, respectively. These figures were obtained by tuning the laser frequency using the PZT of the ECDL, and by recording the waveform of the modulation transfer signal with a digital oscilloscope (see Fig. \ref{figure1}). A continuous frequency scan across a full spectral range of the hyperfine structure ($\sim1$ GHz) could be realized without mode-hopping.

When the ground state of molecular iodine has an odd rotational quantum number, the rovibrational energy level is split into 21 sublevels, which results in 21 hyperfine components. In the observed $R(37)16-1$ and $P(33)16-1$ transitions (see Figs. \ref{R37} and \ref{P33}), all the 21 components were isolated from each other. In the $R(101)17-1$ transition (see Fig. \ref{R101}), the $a_{5}$ and $a_{6}$ components were overlapped. The two weak lines in Fig. \ref{R101} belong to the $P(131)18-1$ transition. The $a_{3}$ component of the $P(131)18-1$ transition could not be clearly assigned from the spectrum in Fig. \ref{P131}. This is considered to be due to the fact that this component overlaps the $a_{15}$ component of the $R(101)17-1$ transition (see Sect. \ref{theoreticalanasec}). The signal-to-noise (SN) ratio of e.g., the $a_{4}$ component of the $R(37)16-1$ transition was about 100 in a bandwidth of 10 Hz. The spectral linewidth of this transition was approximately 1 MHz. The SN ratio and the spectral linewidth are similar to those of our previous experiment \cite{Hong2009}. The frequency stability is estimated to be 10 kHz for a 0.1-s averaging time from the SN ratio and the linewidth. This corresponds to a fractional frequency stability of $2\times10^{-11}$ at a 0.1-s averaging time.


\subsection{Frequency stability}
\label{frequencystabilitysection}
\begin{figure}[t]
\begin{center}
\includegraphics[width=8.5cm,bb=0 50 1010 716]{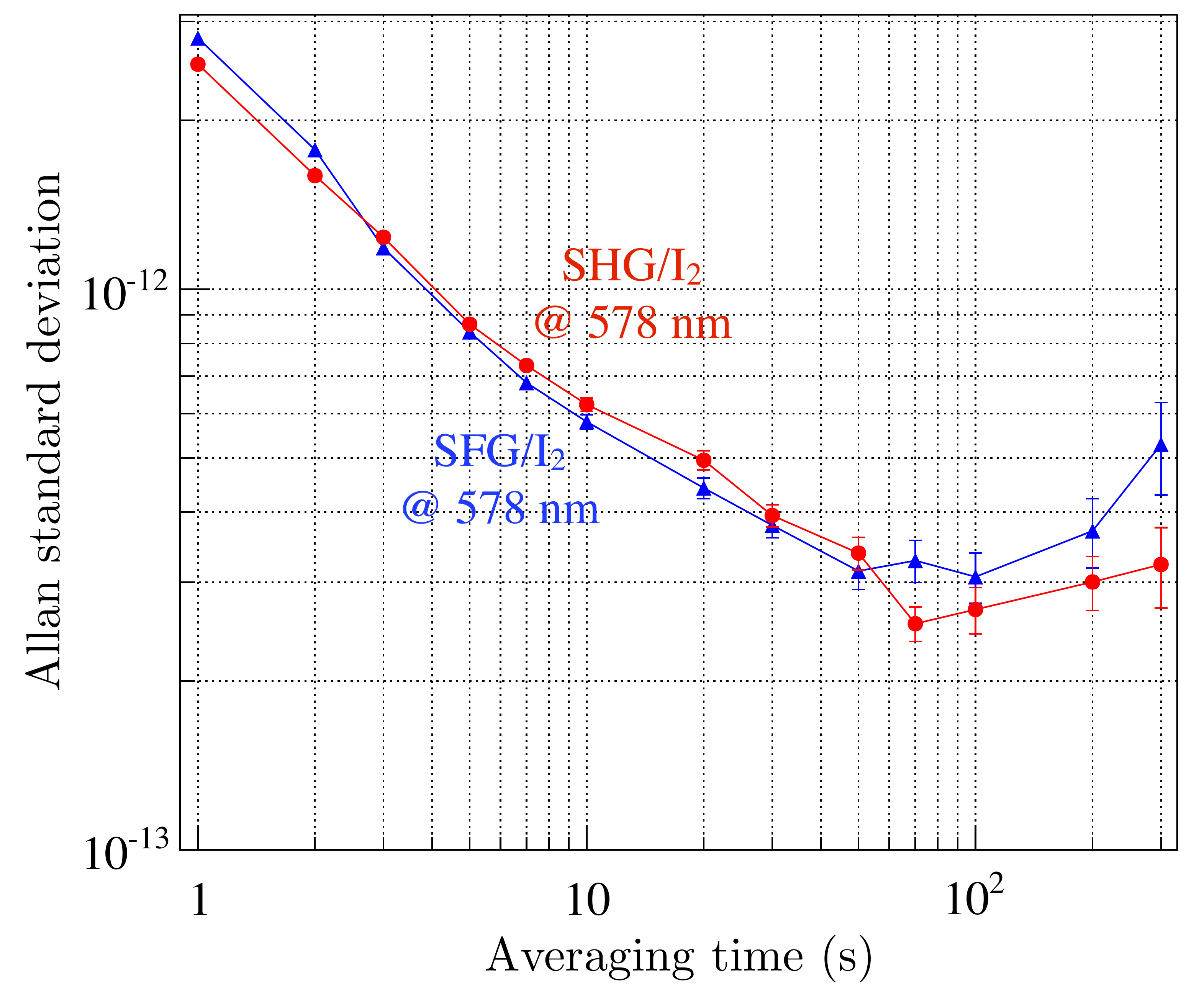}
\end{center}
\caption{Allan standard deviation calculated from the measured beat frequency between the I$_{2}$-stabilized SHG light at 578 nm and the optical frequency comb stabilized to the H-maser (solid red dots). For comparison, the Allan deviation calculated from the measured beat frequency between the I$_{2}$-stabilized SFG light at 578 nm \cite{Hong2009} and the frequency comb stabilized to the H-maser is also shown (solid blue triangles). Both the SHG and SFG light sources were locked on the $a_{4}$ component of the $R(37)16-1$ transition.}
\label{figure2}
\end{figure}

The observed modulation transfer signals were used for frequency stabilization. Figure \ref{figure2} shows the Allan standard deviation calculated from the measured beat frequency between the 578-nm SHG light source stabilized to the $a_{4}$ component of the $R(37)$16-1 transition and the optical frequency comb locked on the H-maser (solid red dots). The frequency counter used in the measurement was a $\Pi$-type counter \cite{Benkler2015} with zero dead time. The observed stability of the beat frequency was mainly limited by the iodine-stabilized SHG light source, because the Allan deviation of UTC(NMIJ) starts from the $10^{-13}$ level at 1 s \cite{Inaba2006}. The frequency stability of the SHG light source was $2.5\times 10^{-12}$ at an averaging time of $\tau=1$ s, and reached $2.5\times10^{-13}$ at $\tau=70$ s. The long-term stability for $\tau>70$ s was limited by a flicker floor at $2.5\times10^{-13}$. At $\tau=1$ s, the stability is basically determined by the SN ratio and the spectral linewidth. Assuming the $1/\tau^{1/2}$ slope, the observed stability is comparable to the estimated frequency stability for the $R(37)16-1:a_{4}$ transition (see Sect. \ref{observationhyperfinesect}). For comparison, the Allan deviation of our solid-state-laser-based SFG light source (see Sect. \ref{introduction}) \cite{Hong2009} locked on the same $R(37)16-1:a_{4}$ transition is also shown in Fig. \ref{figure2} (solid blue triangles). The frequency stability of the present SHG light source was nearly identical to that of the previous SFG light source. This is because the spectral linewidth ($\sim$1 MHz), which determines the frequency stability, is not mainly contributed from the linewidth of the ECDL, but from pressure and power broadening effects. 

\subsection{Absolute frequency measurements}
\label{absfrequencysect}
\begin{figure}[t]
\begin{center}
\includegraphics[width=10cm,bb=0 50 1210 716]{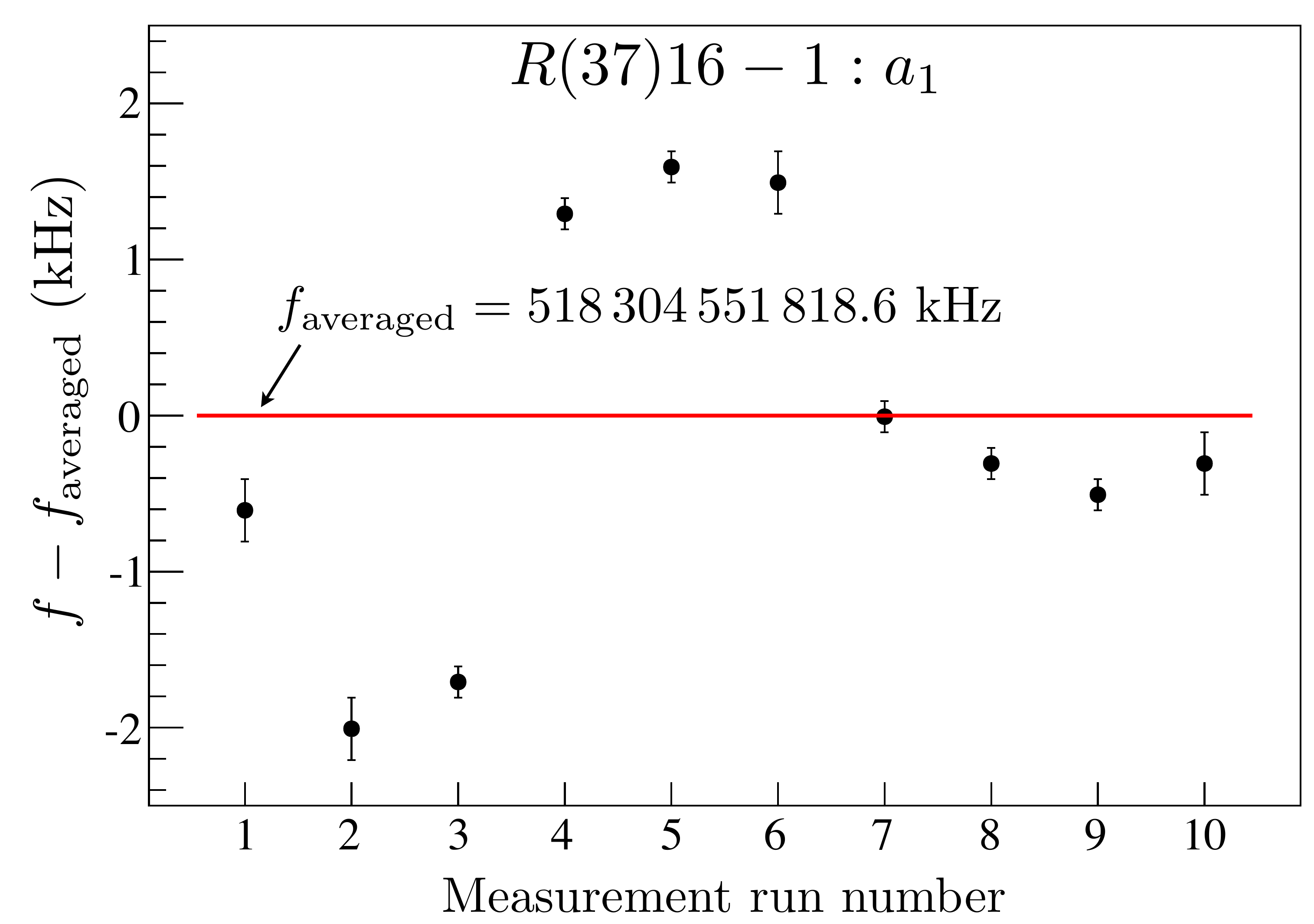}
\end{center}
\caption{Measured absolute frequency of a 578-nm SHG light source locked on the $a_{1}$ component of the $R(37)16-1$ transition. The solid red line indicates the average of the ten measured frequencies.}
\label{repeatability}
\end{figure}

Since the frequency comb is reference to UTC(NMIJ), i.e., the national frequency standard, the absolute frequency of the I$_{2}$-stabilized SHG light source is calculated from the results of the beat measurement. Figure \ref{repeatability} shows the results of ten measurements for the $a_{1}$ component of the $R(37)16-1$ transition obtained over several weeks. During the ten measurements, we repeated the process of locking and unlocking the laser. The servo electronics offset, i.e., a dc voltage offset between the baseline of the spectrum and the lock point, was adjusted to $<1$ mV to avoid the offset in the laser frequency. Each measurement in Fig. \ref{repeatability} was calculated from about 1000 beat frequency data, where each frequency datum was measured by a frequency counter with a gate time of 1 s. The uncertainty bar in this figure was given by the Allan deviation at the longest averaging time. The average of the ten measured frequencies in Fig. \ref{repeatability} was $518\,304\,551\,818.6$ kHz. The standard deviation of the ten measured frequencies was 1.3 kHz, which indicates the repeatability of our measurement. The repeatability is considered to be limited by the long-term frequency drift. The accuracy of UTC(NMIJ) was confirmed to be better than $1\times10^{-13}$, corresponding to an absolute uncertainty of 0.05 kHz in the iodine transition frequency.  

\begin{table}[h]
\caption{Investigated systematic frequency shifts of a 578-nm SHG light source locked on the $a_{1}$ component of the $R(37)16-1$ transition and their uncertainties.}  
	\label{uncertaintyesti}
	\begin{center} 
\begin{tabular}{lll}
\hline
Effect & Sensitivity & Uncertainty\\
\hline
Pressure shift & $-7.7$ kHz/Pa & $<0.6$ kHz \\
Power shift & $-0.8$ kHz/mW & $<0.4$ kHz \\
Modulation phase adjustment & & $<0.5$ kHz \\
Cell impurity & & $5$ kHz\\
\hline
\end{tabular}
\end{center}
\end{table}

\begin{figure}[h]
\begin{center}
\includegraphics[width=11.5cm,bb=0 80 1210 716]{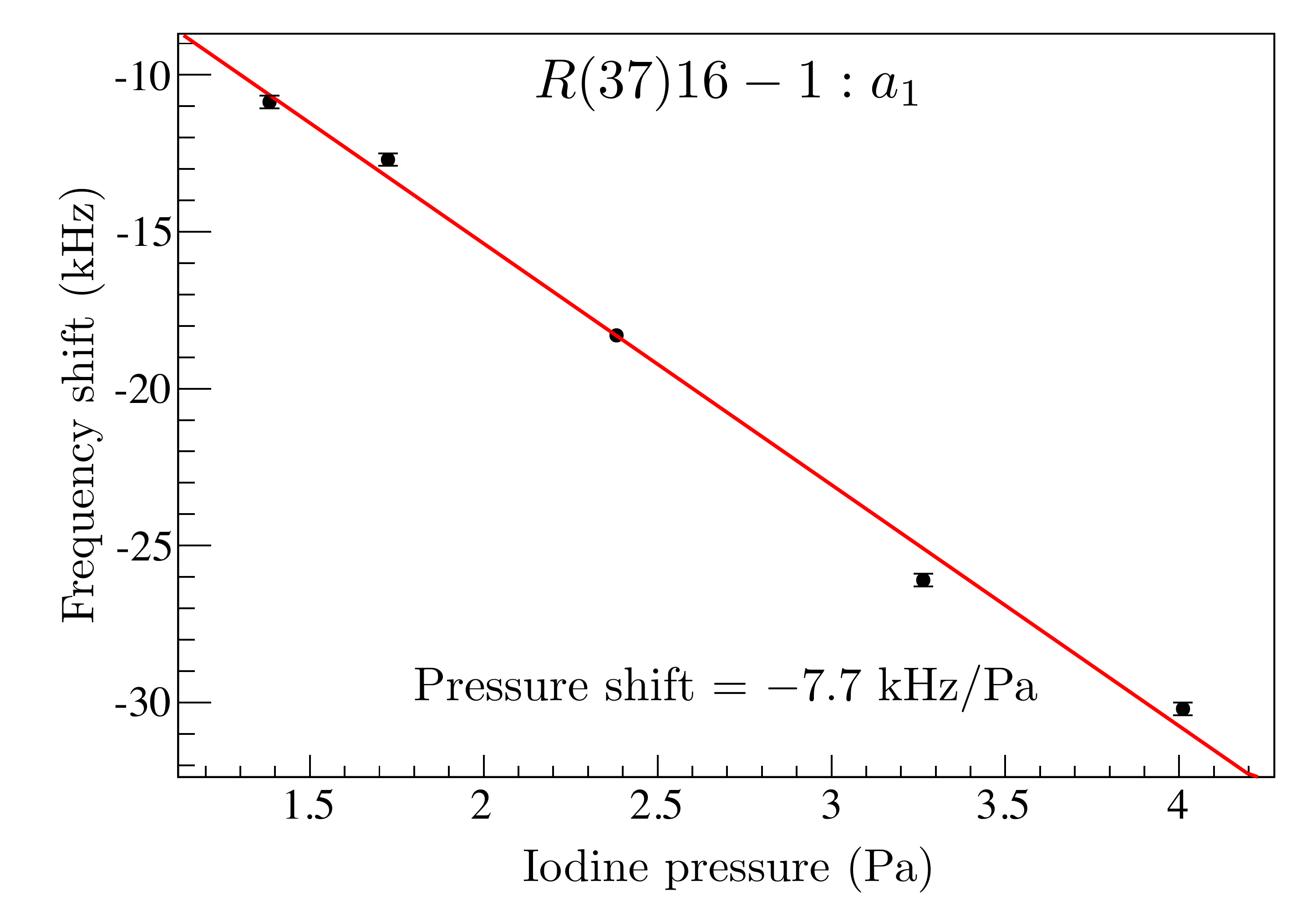}
\end{center}
\caption{Pressure shift of a 578-nm SHG light source locked on the $a_{1}$ component of the $R(37)16-1$ transition. The powers of the pump and probe beams were fixed at $P_{\mathrm{pump}}=4.8$ mW and $P_{\mathrm{probe}}=0.5$ mW, respectively. The solid red line indicates the best fit of a linear function using the weighted least square method.} 
\label{pressureshift}
\end{figure}

\begin{figure}[h]
\begin{center}
\includegraphics[width=11.5cm,bb=0 80 1210 716]{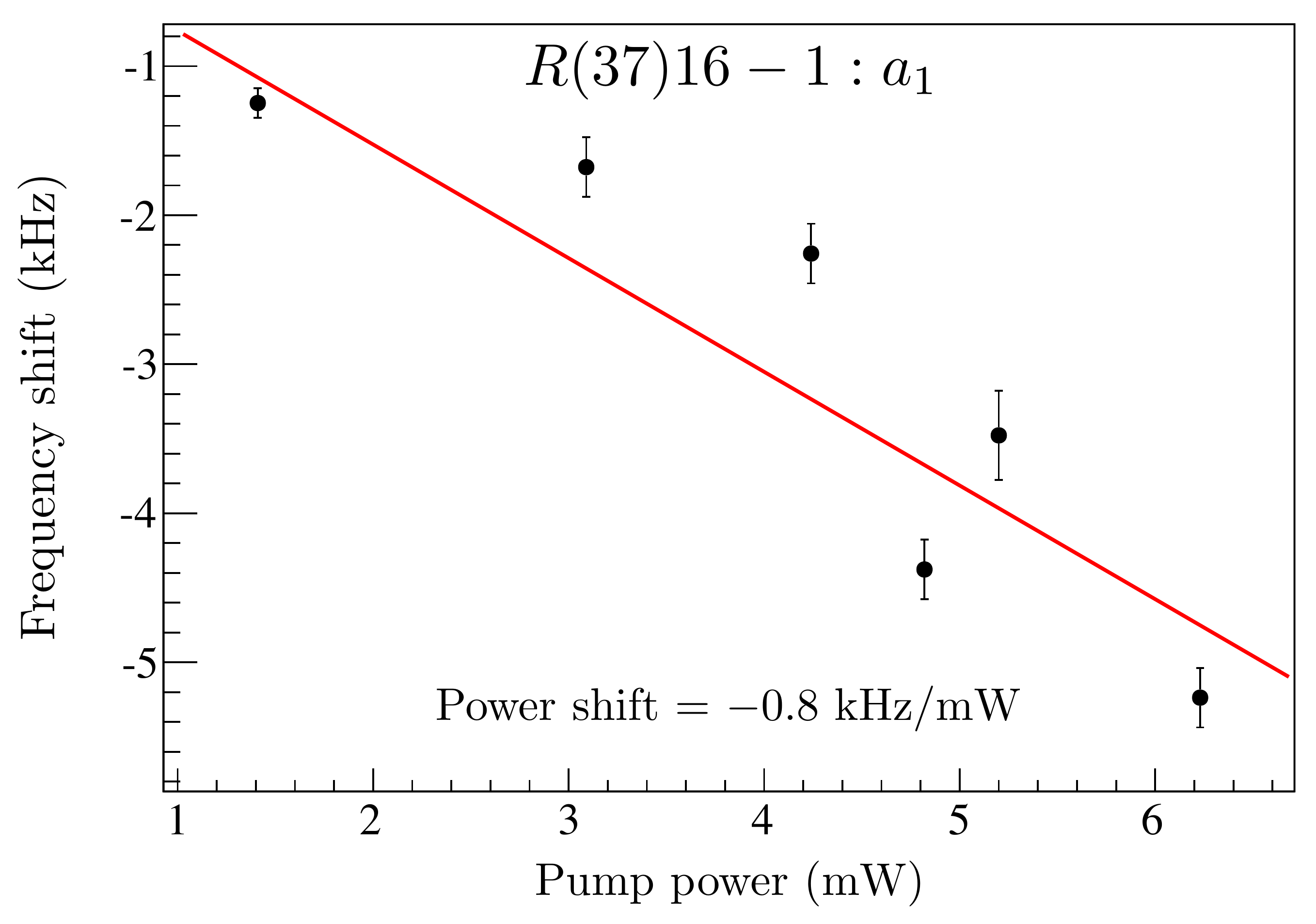}
\end{center}
\caption{Power shift of a 578-nm SHG light source locked on the $a_{1}$ component of the $R(37)16-1$ transition. The cold-finger temperature of the iodine cell was kept at $T=-2$ $^{\circ}$C ($p=3.3$ Pa). The solid red line indicates the best fit of a linear function using the weighted least square method.}
\label{powershift}
\end{figure}

Several systematic frequency shifts of the I$_{2}$-stabilized SHG light source were investigated. Figure \ref{pressureshift} shows the measured pressure shift of the SHG light source locked on the $R(37)16-1:a_{1}$ transition. The measured slope of the pressure shift was $-7.7$ kHz/Pa. The stability of the cold-finger temperature was better than 10 mK. However, it is difficult to determine the temperature of the solid-state iodine in the cold finger, which determines the iodine pressure, since there is a temperature slope between the solid-state iodine and the temperature sensor \cite{Hong2004optcom}. We set the uncertainty in the determination of the temperature of the solid-state iodine at $<0.5$ K. This corresponds to a pressure uncertainty of $<0.08$ Pa, which results in a frequency uncertainty of $<0.6$ kHz. Figure \ref{powershift} shows the measured power shift of the SHG light source locked on the $R(37)16-1:a_{1}$ transition. The measured slope of the power shift was $-0.8$ kHz/mW. The uncertainty in the determination of the laser power was conservatively estimated to be $<10\%$. This gives rise to a frequency uncertainty of $<0.4$ kHz. The misalignment of the pump and probe beams induces a frequency shift \cite{Hong2004optcom}. We checked this effect by varying the pointing direction of the pump beam between $\Delta \theta = 0$ and $3$ mrad. No significant frequency shifts were observed until $\Delta\theta\sim 3$ mrad. This indicates that the frequency shift caused by possible misalignment of the pump and probe beams during the measurements is negligibly small. Our previous experiments \cite{Hong2004optcom} have shown that a frequency shift is caused by modulation phase adjustment, i.e., the adjustment of a phase shifter (see Fig. \ref{figure1}), which shifts the phase between the modulation applied to the pump beam and a local oscillation port of a demodulation device, and the uncertainty resulting from this effect is $<0.35$ kHz. Here we conservatively set the uncertainty resulting from the modulation phase adjustment at $<0.5$ kHz. Contamination in the iodine cell causes a frequency shift. International comparisons have shown that this effect can add an uncertainty of 5 kHz to measurement results \cite{Eickhoff1995}. 


The investigated systematic frequency shifts of the I$_{2}$-stabilized light source and their uncertainties are summarized in Table \ref{uncertaintyesti}. However, as discussed in Sect. \ref{discussionandconclusion}, there may be some unknown systematic uncertainty facts in the experiment. The total uncertainty of the measured absolute frequencies is estimated in Sect. \ref{discussionandconclusion}.

\subsection{Measurements of hyperfine splittings}
\label{hyeperfinesecst}

\begin{table}[t]
\caption{Measured absolute frequencies of the $a_{1}$ component of the four observed rovibrational transitions. The uncertainties are estimated in Sect. \ref{discussionandconclusion}.}  
	\label{absoluteresultcomp}
	\begin{center} 
\begin{tabular}{lcc}
\hline
Transition & Component & Absolute frequency (kHz)\\
\hline
$R(37)16-1$ & $a_{1}$ & $518\,304\,551\,819(7)$\\
$P(33)16-1$ & $a_{1}$ & $518\,287\,542\,790(7)$\\
$R(101)17-1$ & $a_{1}$ & $518\,288\,804\,784(7)$\\
$P(131)18-1$ & $a_{1}$ & $518\,289\,467\,712(7)$\\
\hline
\end{tabular}
\end{center}
\end{table}

\begin{table}[h]
\caption{Observed and calculated hyperfine splittings of the $R(37)16-1$ transition$^{a}$.}  
	\label{R37table}
	\begin{center} 
\begin{tabular}{crrrccc}
\hline
Hyperfine & Obs. & Calc. & Obs. $-$ Calc. & & &\\
Components & (kHz) & (kHz) & (kHz) & Weight & $F^{'}$ &$I$\\
\hline
$a_{1}$ & 0 & 0.8 & $-0.8$ & 1.0 & 33 & 5\\
$a_{2}$ & 28196.1 & 28196.6 & $-0.5$ & 1.0 & 38 & 1\\
$a_{3}$ & 54556.3 & 54556.7 & $-0.4$ & 1.0 & 43 & 5\\
$a_{4}$ & 263163.9 & 263163.9 & $0.0$ & 1.0 & 34 & 5\\
$a_{5}$ & 311346.4 & 311345.5 & $0.9$ & 1.0 & 39 & 3\\ 
$a_{6}$ & 324920.7 & 324921.2 & $-0.5$ & 1.0 & 37 & 3\\
$a_{7}$ & 370347.2 & 370345.8 & $1.4$ & 1.0 & 42 & 5\\
$a_{8}$ & 433825.8 & 433825.1 & 0.7 & 1.0 & 35 & 5 \\
$a_{9}$ & 457082.9 & 457081.2 & 1.7 & 1.0 & 36 & 3\\
$a_{10}$ & 467068.4 & 467070.7 & $-2.3$ & 1.0 & 40 & 3\\
$a_{11}$ & 480842.2 & 480841.7 & 0.5 & 1.0 & 41 & 3\\
$a_{12}$ & 571555.6 & 571553.9 & $1.7$ & 1.0 & 35 & 3\\
$a_{13}$ & 607525.6 & 607526.6 & $-1.0$ & 1.0 & 38 & 3\\
$a_{14}$ & 651329.6 & 651330.3 & $-0.7$ & 1.0 & 41 & 5\\
$a_{15}$ & 705758.5 & 705760.8 & $-2.3$ & 1.0 & 36 & 5\\
$a_{16}$ & 738305.0 & 738304.3 & 0.7 & 1.0 & 37 & 5\\
$a_{17}$ & 762328.3 & 762328.6 & $-0.3$ & 1.0 & 39 & 5\\
$a_{18}$ & 795124.9 & 795125.6 & $-0.7$ & 1.0 & 40 & 5\\
$a_{19}$ & 879552.4 & 879553.8 & $-1.4$ & 1.0 & 37 & 1\\
$a_{20}$ & 893838.3 & 893836.4 & 1.9 & 1.0 & 38 & 5\\
$a_{21}$ & 914090.3 & 914090.0 & 0.3 & 1.0 & 39 & 1\\
\hline
\end{tabular}
\end{center}
$^{a}$ The standard deviation of the fit is 1.4 kHz. 
\end{table}

\begin{table}[h]
\caption{Observed and calculated hyperfine splittings of the $P(33)16-1$ transition$^{a}$.}  
	\label{P33table}
	\begin{center} 
\begin{tabular}{crrrccc}
\hline
Hyperfine & Obs. & Calc. & Obs. $-$ Calc. & & &\\
Components & (kHz) & (kHz) & (kHz) & Weight & $F^{'}$ & $I$\\
\hline
$a_{1}$ & 0 & $-0.4$ & $0.4$ & 1.0 & 27 & 5\\
$a_{2}$ & 29346.9 & 29347.3 & $-0.4$ & 1.0 & 32 & 1\\
$a_{3}$ & 56578.7 & 56581.5 & $-2.8$ & 1.0 & 37 & 5\\
$a_{4}$ & 259954.3 & 259954.3 & $0.0$ & 1.0 & 28 & 5\\
$a_{5}$ & 311400.3 & 311400.1 & $0.2$ & 1.0 & 33 & 3\\ 
$a_{6}$ & 327088.6 & 327089.3 & $-0.7$ & 1.0 & 31 & 3\\
$a_{7}$ & 375292.3 & 375294.7 & $-2.4$ & 1.0 & 36 & 5\\
$a_{8}$ & 432137.2 & 432137.2 & $0.0$ & 1.0 & 29 & 5\\
$a_{9}$ & 458103.2 & 458099.5 & 3.7 & 0.0 & 30 & 3\\
$a_{10}$ & 467776.7 & 467780.9 & $-4.2$ & 0.0 & 34 & 3\\
$a_{11}$ & 482269.0 & 482269.6 & $-0.6$ & 1.0 & 35 & 3\\
$a_{12}$ & 570547.2 & 570544.5 & $2.7$ & 1.0 & 29 & 3\\
$a_{13}$ & 608476.8 & 608474.9 & $1.9$ & 1.0 & 32 & 3\\
$a_{14}$ & 655962.8 & 655961.0 & $1.8$ & 1.0 & 35 & 5\\
$a_{15}$ & 702467.4 & 702468.0 & $-0.6$ & 1.0 & 30 & 5\\
$a_{16}$ & 737777.9 & 737776.9 & 1.0 & 1.0 & 31 & 5\\
$a_{17}$ & 763752.0 & 763750.3 & $1.7$ & 1.0 & 33 & 5\\
$a_{18}$ & 799485.7 & 799485.7 & $0.0$ & 1.0 & 34 & 5\\
$a_{19}$ & 879016.8 & 879019.7 & $-2.9$ & 1.0 & 31 & 1\\
$a_{20}$ & 893989.8 & 893993.1 & $-3.3$ & 1.0 & 32 & 5\\
$a_{21}$ & 916193.1 & 916192.9 & 0.2 & 1.0 & 33 & 1\\
\hline
\end{tabular}
\end{center}
$^{a}$The standard deviation of the fit is 2.0 kHz. 
\end{table}

\begin{table}[h]
\caption{Observed and calculated hyperfine splittings of the $R(101)17-1$ transition$^{a}$.}  
	\label{R101table}
	\begin{center} 
\begin{tabular}{crrrccc}
\hline
Hyperfine & Obs. & Calc. & Obs. $-$ Calc. & & & \\
Components & (kHz) & (kHz) & (kHz) & Weight & $F^{'}$ & $I$\\ 
\hline
$a_{1}$ & 0 & $-0.2$ & 0.2 & 1.0 & 97 & 5 \\
$a_{2}$ & 27454.1 & 27454.1 & $0.0$ & 1.0 & 102 & 1\\
$a_{3}$ & 54174.2 & 54173.2 & $1.0$ & 1.0 & 107 & 5\\
$a_{4}$ & 283539.3 & 283539.2 & $0.1$ & 1.0 & 98 & 5\\
$a_{5}$ & $-$ & 316374.8 & $-$ & 0.0 & 101 & 3\\ 
$a_{6}$ & $-$ & 318200.2& $-$ & 0.0 & 103 & 3\\
$a_{7}$ & 350221.7 & 350222.7 & $-1.0$ & 1.0 & 106 & 5\\
$a_{8}$ & 442770.7 & 442771.3 & $-0.6$ & 1.0 & 99 & 5\\
$a_{9}$ & 453446.2 & 453448.8 & $-2.6$ & 1.0 & 100 & 3\\
$a_{10}$ & 470642.4 & 470641.2 & $1.2$ & 1.0 & 104 & 3\\
$a_{11}$ & 479924.2 & 479926.7 & $-2.5$ & 1.0 & 105 & 3\\
$a_{12}$ & 582310.6 & 582310.7 & $-0.1$ & 1.0 & 99 & 3\\
$a_{13}$ & 607068.5 & 607067.6 & $0.9$ & 1.0 & 102 & 3\\
$a_{14}$ & 632708.9 & 632708.4 & $0.5$ & 1.0 & 105 & 5\\
$a_{15}$ & 728165.0 & 728234.3 & $-69.3$ & 0.0 & 100 & 5\\
$a_{16}$ & 743986.8 & 743986.6 & 0.2 & 1.0 & 101 & 5\\
$a_{17}$ & 759392.9 & 759391.9 & $1.0$ & 1.0 & 103 & 5\\
$a_{18}$ & 774863.0 & 774862.3& $0.7$ & 1.0 & 104 & 5\\
$a_{19}$ & 886914.6 & 886912.7  & $1.9$ & 1.0 & 101 & 1\\
$a_{20}$ & 896283.9 & 896283.5 & $0.4$ & 1.0 & 102 & 5\\
$a_{21}$ & 906694.8 & 906698.2 & $-3.4$ & 1.0 & 103 & 1\\
\hline
\end{tabular}
\end{center}
$^{a}$The standard deviation of the fit is 1.7 kHz. 
\end{table}

\begin{table}[h]
\caption{Observed and calculated hyperfine splittings of the $P(131)18-1$ transition$^{a}$.}  
	\label{P131table}
	\begin{center} 
\begin{tabular}{crrrccc}
\hline
Hyperfine & Obs. & Calc. & Obs. $-$ Calc. & & & \\ 
Components & (kHz) & (kHz) & (kHz) & Weight & $F^{'}$ & $I$\\
\hline
$a_{1}$ & 0 & 0.3 & $-0.3$ & 1.0 & 125 & 5\\
$a_{2}$ & 32859.0 & 32858.9 & $0.1$ & 1.0 & 130 & 1\\ 
$a_{3}$ & $-$ & 65014.9 & $-$ & 0.0 & 135 & 5\\
$a_{4}$ & 287115.8 & 287114.9 & $0.9$ & 1.0 & 126 & 5\\
$a_{5}$ & 319406.7 & 319398.2 & $8.5$ & 0.0 & 129 & 3\\ 
$a_{6}$ & 325476.5 & 325484.0 & $-7.5$ & 0.0 & 131 & 3\\
$a_{7}$ & 357046.3 & 357047.3 & $-1.0$ & 1.0 & 134 & 5\\
$a_{8}$ & 445414.1 & 445411.9 & $2.2$ & 1.0 & 127 & 3\\
$a_{9}$ & 455840.0 & 455843.3 & $-3.3$ & 1.0 & 128 & 3\\
$a_{10}$ & 478410.9 & 478406.7 & $4.2$ & 1.0 & 132 & 3\\
$a_{11}$ & 487963.0 & 487963.3 & $-0.3$ & 1.0 & 133 & 3\\
$a_{12}$ & 585824.3 & 585826.4 & $-2.1$ & 1.0 & 127 & 5\\
$a_{13}$ & 612001.0 & 612002.7 & $-1.7$ & 1.0 & 130 & 3\\
$a_{14}$ & 638605.8 & 638602.8  & $3.0$ & 1.0 & 133 & 5\\
$a_{15}$ & 733868.8 & 733867.8 & $1.0$ & 1.0 & 128 & 5\\
$a_{16}$ & 748367.9 & 748367.0 & $0.9$ & 1.0 & 129 & 5\\
$a_{17}$ & 764869.2 & 764868.2 & $1.0$ & 1.0 & 131 & 5\\
$a_{18}$ & 779039.9 & 779040.3 & $-0.4$ & 1.0 & 132 & 5\\
$a_{19}$ & 891585.6 & 891584.0 & $1.6$ & 1.0 & 129 & 1\\
$a_{20}$ & 901153.6 & 901154.4 & $-0.8$ & 1.0 & 130 & 5\\
$a_{21}$ & 911468.9 & 911471.2  & $-2.3$ & 1.0 & 131 & 1\\
\hline
\end{tabular}
\end{center}
$^{a}$The standard deviation of the fit is 2.2 kHz. 
\end{table}
\begin{table*}[t]
\caption{Fitted hyperfine constants of the observed transitions.}  
	\label{fitconstanttable}
	\begin{center} 
\begin{tabular}{ccccc}
\hline
Constant & $R(37)16-1$ & $P(33)16-1$ & $R(101)17-1$ & $P(131)18-1$\\
\hline
$\Delta e Qq$ (kHz) & 1935034.2(2.0) & 1935058.6(2.9) & $1933277.0(2.0)$ & $1931596(3)$\\
$\Delta C$ (kHz) & 35.3759(29) & 35.334(5) & $38.1581(13)$ & $41.0014(16)$\\
$\Delta d$ (kHz) & $-16.85(14)$ & $-16.83(20)$ & $-17.82(21)$ & $-19.45(20)$\\
$\Delta \delta$ (kHz) & $-4.21(11)$ & $-4.05(16)$ & $-4.51(19)$ & $-4.47(14)$\\
\hline
\end{tabular}
\end{center}
\end{table*}

To measure the hyperfine splittings, the SHG light source was stabilized in succession to all the observed hyperfine components of the four rovibrational transitions (see Figs. \ref{R37}, \ref{P33}, \ref{R101}, and \ref{P131}). An absolute frequency measurement was carried out for each hyperfine component. Table \ref{absoluteresultcomp} shows the measured absolute frequency of the $a_{1}$ component of each observed rovibrational transition. Tables \ref{R37table}, \ref{P33table}, \ref{R101table}, and \ref{P131table} list the hyperfine splittings obtained by calculating the frequency difference between the measured absolute frequency of each component and that of the $a_{1}$ component. Each obtained result in these tables is an average of about 1000 values of the beat frequency data, where each datum was measured by a frequency counter with a gate time of 1 s. The measurement uncertainty was estimated to be $1-2$ kHz from the repeatability of the I$_{2}$-stabilized light source (see Sect. \ref{absfrequencysect}).

\section{Theoretical analysis}
\label{theoreticalanasec}
The measured hyperfine splittings were used to determine the hyperfine constants of the four observed transitions. The hyperfine interactions of I$_{2}$ can be described by an effective Hamiltonian $H_{\mathrm{HFS}}$ \cite{Borde1981},
\begin{equation}
H_{\mathrm{HFS}} = eQq\cdot H_{\mathrm{EQ}} + C \cdot H_{\mathrm{SR}} + d \cdot H_{\mathrm{TSS}} + \delta \cdot H_{\mathrm{SSS}},
\label{fourthtermhamilton}
\end{equation}
where $H_{\mathrm{EQ}}$, $H_{\mathrm{SR}}$, $H_{\mathrm{TSS}}$, and $H_{\mathrm{SSS}}$ respectively, represent the electric quadrupole, spin-rotation, tensor spin-spin, and scalar spin-spin interactions, and $eQq$, $C$, $d$, and $\delta$ denote the corresponding hyperfine constants for each of those interactions. This four-term effective Hamiltonian has been widely used to analyze experimental hyperfine spectra of I$_{2}$. We calculated the eigenstates of the Hamiltonian based on the procedure described by Bord\'e $et$ $al$ \cite{Borde1981}. To include the effect of the couplings between different rotational states induced by the electric quadrupole interactions, we used a basis set containing not only the rotational state $J$ but also the states $J\pm2$ and $J\pm4$. In this calculation, the rotational Hamiltonian $H_{\mathrm{R}}$,
\begin{equation}
\Braket{JIF|H_{\mathrm{R}}|JIF}=BJ(J+1)-DJ^{2}(J+1)^{2}+HJ^{3}(J+1)^{3},
\end{equation}
was introduced, where $I$ denotes the total nuclear spin and $F$ the total angular momentum. The rotational constants $B$, $D$, and $H$ were taken from Ref. \cite{Gerstenkorn1985}. Since the selection rules of the main transitions between the ground and excited states are strict ($\Delta F = \Delta J$, $\Delta J = \pm1$), the hyperfine splitting patterns are similar to those of the ground or excited states, except that the splittings are scaled by the differences between the hyperfine constants in the ground and excited states, i.e., 
\begin{eqnarray}
\Delta eQq &=& eQq^{'} - eQq^{''},\nonumber\\
\Delta C &=& C^{'} - C^{''}, \nonumber\\
\Delta d &=& d^{'} - d^{''}, \nonumber\\
\Delta \delta &=& \delta^{'} - \delta^{''}.
\end{eqnarray}
By theoretically fitting the observed hyperfine spectra, we can obtain accurate values for $\Delta eQq$, $\Delta C$, $\Delta d$, and $\Delta \delta$ but only a very crude estimate of the absolute values of the ground or excited-state hyperfine constants. 

In the present calculation, a nonlinear least-square fit of the calculated hyperfine splittings to the measured values was carried out using the ROOT analysis software of CERN \cite{Brun1997}. In this fit, the hyperfine constants in the excited state $eQq^{'}$, $C^{'}$, $d^{'}$, and $\delta^{'}$ were treated as free parameters. The ground-state constant $eQq^{''}$ was tentatively estimated using a formula describing the $J$ dependence of $eQq^{''}$ with $v^{''}=0$ \cite{Hong2001JOSAB379}. The other ground-state constants $C^{''}$, $d^{''}$, and $\delta^{''}$ were fixed to the values for $v^{''}=0$ given in Ref. \cite{Wallerand1999}. The choice of these fixed parameters may not be suitable for the ground state with $v^{''}=1$ involved in the presently observed transitions, but it has negligible effect on the determination of the differences in the constants $\Delta eQq$, $\Delta C$, $\Delta d$, and $\Delta \delta$, as described above.

The calculated hyperfine splittings and their differences from the observed values are listed in Tables \ref{R37table}, \ref{P33table}, \ref{R101table}, and \ref{P131table}. In these tables, the weight ranges from 0.0 to 1.0. A weight of 0.0 corresponds to no weight in the fit. The agreement between the experimental and theoretical values is comparable to a measurement uncertainty of $1-2$ kHz (see Sect. \ref{hyeperfinesecst}). The standard deviations of the fit for the four observed transitions were $1-2$ kHz. For the $P(33)16-1$ transition (see Table \ref{P33table}), the $a_{9}$ and $a_{10}$ components were excluded from the calculation (weight = 0.0), since they are too close to each other. For the $R(101)17-1$ transition (see Table \ref{R101table}), the $a_{15}$ component was excluded due to the large difference between the observed and calculated values. We believe this component is affected by the $a_{3}$ component of the $P(131)18-1$ transition, which could not be assigned from the sub-Doppler spectrum (see Fig. \ref{P131}). The frequency difference between these components is expected to be only $\sim$200 kHz, which is smaller than the spectral linewidth ($\sim$1 MHz). The absolute frequency of the $P(131)18-1:a_{3}$ transition was calculated from the measured absolute frequency of the $P(131)18-1:a_{1}$ transition (see Table \ref{absoluteresultcomp}) and the calculated hyperfine splitting of the $P(131)18-1:a_{3}$ transition (see Table \ref{P131table}). For the $P(131)18-1$ transition (see Table \ref{P131table}), the $a_{5}$ and $a_{6}$ components are too close to each other, and thus were excluded in the calculation (weight = 0.0).

Table \ref{fitconstanttable} shows the fitted hyperfine constants for the four observed transitions. The uncertainties of the electric quadrupole constant $\Delta eQq$ for the four transitions were $2-3$ kHz. The other hyperfine constants $\Delta C$, $\Delta d$, and $\Delta \delta$ were also deduced with high accuracy.

\section{Discussion and conclusion}
\label{discussionandconclusion}
Our previous experiment determined the absolute frequency of a 578-nm SFG light source locked on the $R(37)16-1:a_{1}$ transition as $518\,304\,551\,833(2)$ kHz \cite{Hong2009}. This was obtained with the same iodine cell used in the present experiment and the parameters $T=-5$ $^{\circ}$C ($p=2.4$ Pa), $P_{\mathrm{pump}}=2.1$ mW, $P_{\mathrm{probe}}=0.21$ mW, and $d\sim1.5$ mm. For these parameters, the present experimental data on the pressure and power shifts (see Figs. \ref{pressureshift} and \ref{powershift}) provide the absolute frequency of that transition as $518\,304\,551\,828$ kHz. The difference between the present and previous results is 5 kHz, which is larger than the uncertainty caused by the pressure shift, the power shift, and the modulation phase adjustment as shown in Table \ref{uncertaintyesti}. It should be noted here that the contamination in the iodine cell does not contribute to this difference, since we used the same iodine cell for both experiments and thus the frequency shift due to the cell impurity is cancelled out. This difference implies that some unknown systematic effects shifted the frequency of the  I$_{2}$-stabilized light source. This can be related to the fact that the long-term frequency stability for $\tau>70$ s was limited by the flicker floor (see Sect. \ref{frequencystabilitysection}), and that the repeatability of the absolute frequency measurements was worse than the expected uncertainty of the investigated systematic frequency shifts (see Sect. \ref{absfrequencysect}). Taking these facts into consideration, we included the difference of 5 kHz as an uncertainty resulting from unknown systematic effects. By adding this uncertainty to the uncertainty of the investigated systematic shifts (see Table \ref{uncertaintyesti}), the total uncertainty of the absolute frequency of the I$_{2}$-stabilized light source was estimated to be 7 kHz (relatively $1.4\times10^{-11}$).

\begin{figure}[h]
\begin{center}
\includegraphics[width=10cm,bb=0 100 1210 716]{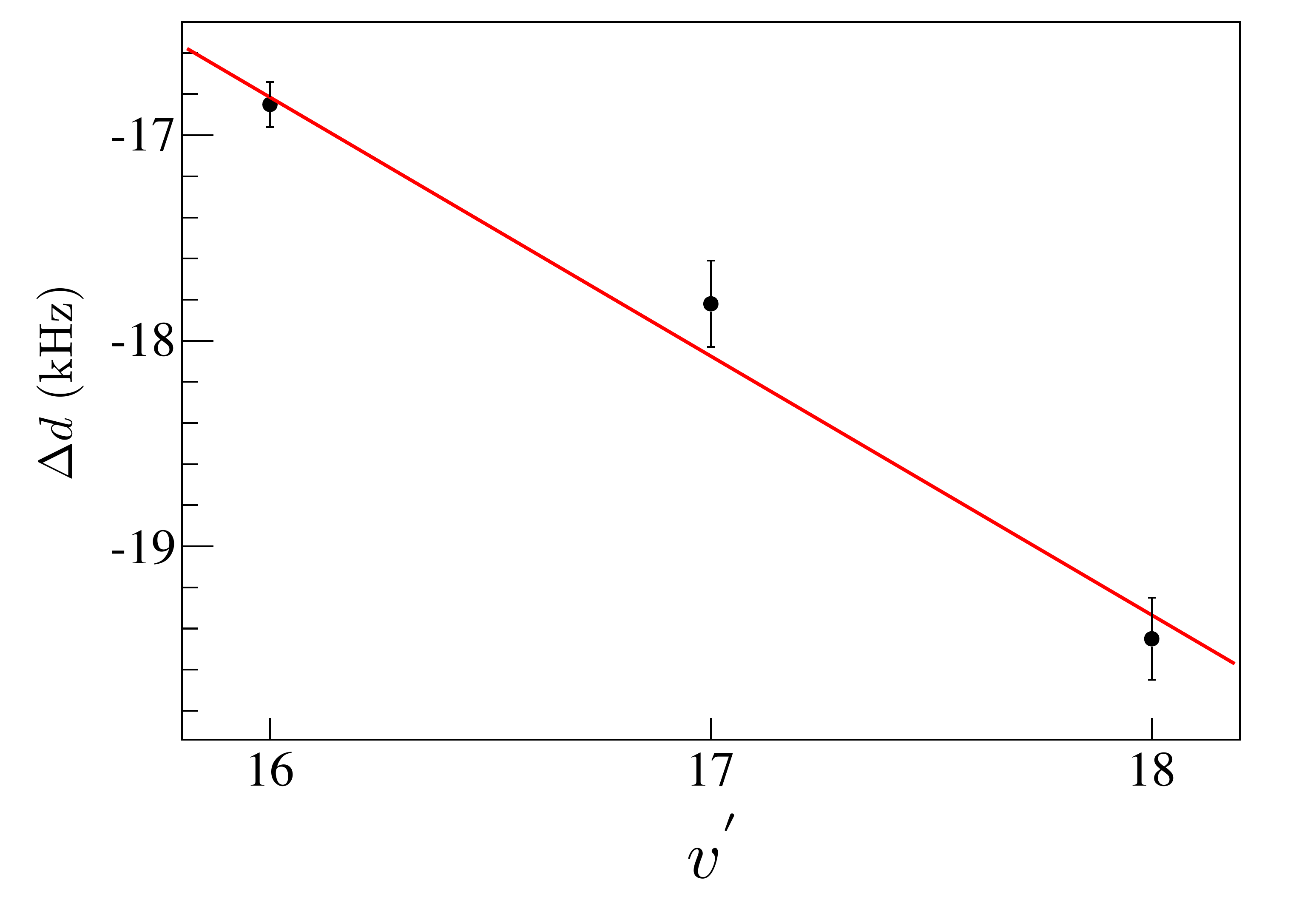}
\end{center}
\caption{Hyperfine constants $\Delta d$ as a function of the vibrational quantum number $v^{'}$. The solid red line shows the best fit of a function including the $(v^{'}+1/2)$ term.} 
\label{scalarspinplot}
\end{figure}

\begin{figure}[h]
\begin{center}
\includegraphics[width=10cm,bb=0 100 1210 716]{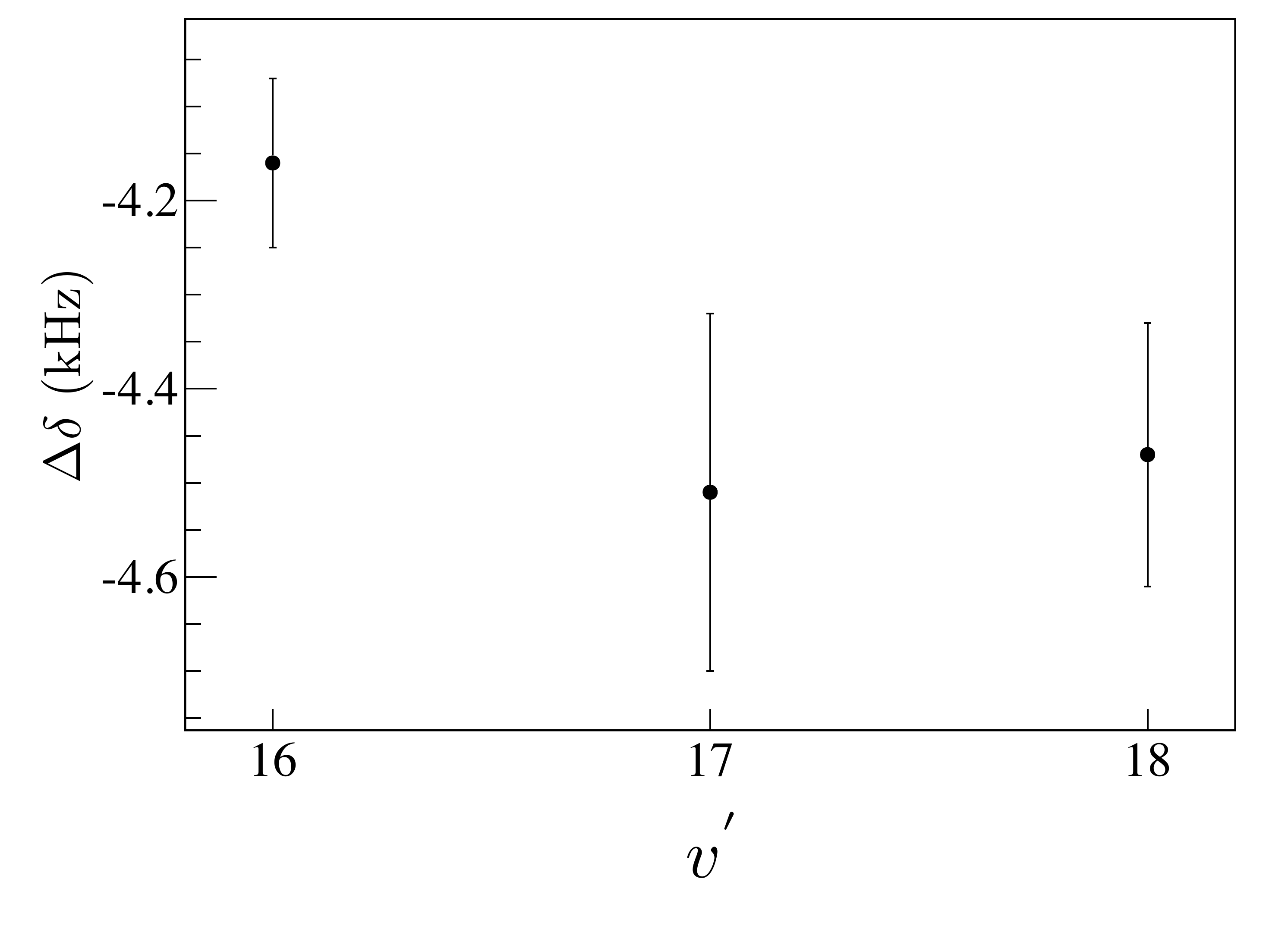}
\end{center}
\caption{Hyperfine constants $\Delta \delta$ as a function of the vibrational quantum number $v^{'}$.}
\label{tensorspinplot}
\end{figure}

Our previous studies \cite{Hong2001JOSAB1416} have shown that the tensor spin-spin and scalar spin-spin constants exhibit no rotation dependence within the measurement uncertainty. On the other hand, these hyperfine constants are dependent on the excited-state vibrational quantum number $v^{'}$ \cite{Hong2002JOSAB,Chen2004}. In the present work, we obtained $\Delta d$ and $\Delta \delta$ for $v^{'}=16-18$ (see Table \ref{fitconstanttable}), whereas in our previous paper we studied the vibrational dependence for $v^{'}=32-43$ \cite{Hong2002JOSAB}. Figures \ref{scalarspinplot} and \ref{tensorspinplot} respectively, show $\Delta d$ and $\Delta \delta$ as a function of $v^{'}$. In these figures, $\Delta d$ and $\Delta \delta$ for $v^{'}=16$ were obtained by calculating the weighted averages of the hyperfine constants determined from the $R(37)16-1$ and $P(33)16-1$ transitions. Vibrational dependence was observed for $\Delta d$, but was not very clear for $\Delta \delta$. We could fit $\Delta d$ to the following formula,
\begin{equation}
\Delta d(v^{'})=4(2)-1.3(1)(v^{'}+1/2)\,\, \mathrm{kHz}.
\end{equation}
The fitting result is shown in Fig. \ref{scalarspinplot} as a solid red line. This formula could be compared with our previous formula for the excited-state constant $d^{'}(v^{'})$ \cite{Hong2002JOSAB}, if the ground-state constant $d^{''}$ for $v^{''}=1$ was available.


A more compact I$_{2}$-stabilized laser system operating at 578 nm would be useful for various applications including Yb atom research (see Sect. \ref{introduction}). We have previously developed a compact I$_{2}$-stabilized laser at 531 nm using a coin-sized laser module, which consists of a 1062 nm diode laser and a PPLN crystal \cite{Kobayashi2015}. All optical parts of the laser system were arranged on a 20 cm $\times$ 30 cm breadboard, which is much smaller than the present experimental setup. This 531-nm laser is currently being used for the interferometric measurement of gauge blocks \cite{Bitou2003}. A similar coin-sized module at 578 nm could in principle be constructed. This module would be attractive for use in the gauge block measurement, since the measurement requires several light sources emitting at different wavelengths \cite{Bitou2004}. 

In conclusion, we have measured the absolute frequencies of a 578-nm light source stabilized to 81 hyperfine components utilizing the SHG of a 1156-nm ECDL and a fiber-based frequency comb. The frequency stability of the I$_{2}$-stabilized SHG light source was $2.5\times10^{-12}$ at $\tau=1$ s, and reached $2.5\times10^{-13}$ at $\tau=70$ s. Table \ref{absoluteresultcomp} summarizes the absolute frequencies of the $a_{1}$ component of the four observed rovibrational transitions. The absolute frequencies of the other hyperfine components are obtained by adding the absolute frequencies of the $a_{1}$ component in Table \ref{absoluteresultcomp} to the hyperfine splittings listed in Tables \ref{R37table}, \ref{P33table}, \ref{R101table}, and \ref{P131table}. The relative uncertainties of the measured absolute frequencies were typically $1.4\times10^{-11}$. Accurate hyperfine constants were deduced by fitting the hyperfine splittings to the four-term Hamiltonian. The observed transitions can provide good frequency references in the 578-nm region. In the future, we plan to extend our measurements of the hyperfine components to the other yellow lines of I$_{2}$ near 578 nm. 

\acknowledgements
We are grateful to M. Amemiya and T. Suzuyama for maintaining UTC(NMIJ). We thank M. Abe, K. Iwakuni, and H. Sasada for discussions regarding the theoretical analysis. Part of this work was supported by Japan Society for the Promotion of Science (JSPS) KAKENHI Grant Number 15H02028.

\end{document}